\documentstyle[aps,prl]{revtex}
 
\input{epsf}
\epsfverbosetrue

\begin{document}
\draft
 
\twocolumn[\hsize\textwidth\columnwidth\hsize\csname
@twocolumnfalse\endcsname

\title{The Chaotic Ball: An Intuitive Analogy for EPR
Experiments\cite{Ref}}

\author{Caroline H Thompson\cite{CHT/email+}\\
        Department of Computer Science, University of Wales,    Aberystwyth, \\
        SY23 3DB, U.K.}
\date{18 November, 1996}
\maketitle

\begin{abstract}
Actual realisations of EPR  experiments do {\em not} demonstrate non-locality.
A model is presented that should enable non-specialists as well as specialists
to understand how easy it is to find realistic explanations for the
observations.  The model also suggests new areas where realistic
(``hidden-variable'') models can give valid predictions whilst quantum
mechanics fails.  It offers straightforward explanations for some anomalies
that Aspect was unable to account for, providing perhaps the first
experimental evidence that a hidden-variable theory can be {\em superior}
to quantum mechanics.  The apparent success of quantum mechanics in predicting
results is shown to be largely due to the use of unjustifiable and biased
analysis of the data.  Data that has been discarded because it did not lead
to a valid Bell's test may give further evidence that hidden variables exist.
\end{abstract}

\vskip2pc]

 
\setcounter{footnote}{1}

\section{Introduction}

This is the first of a set of papers  that look at actual 
Einstein-Podolksy-Rosen (EPR) experiments from the point of view of a
scientifically and statistically literate person
who is not a specialist in quantum theory.  The incentive to study this area
came from a book review that mentioned that experiments at Orsay had
demonstrated that action-at-a-distance really happened.  This is clearly
impossible.  Experience as a practical statistician suggested that if the
assumptions and analysis methods were checked, rational explanations of the
actual observations would be found.

An abundance of papers and books have been written discussing the whole EPR
question.  Just a few give explanations that appear to be entirely realistic.
These few are in broad agreement with the present paper, recognising the 
importance of variable detection probabilities (VDPs) 
\cite{Marshall/Santos/Selleri83,Pascazio88,Pearle70}. 
  Theoretical papers,
for example the report by Clauser and Shimony \cite{Clauser/Shimony78},
emphasise the related problem, that the original thought experiments 
\cite{Bohm51,Einstein/Podolsky/Rosen35} and John Bell's inequalities
in their original forms \cite{Bell64}, assume that every particle produced
is detected.  In real experiments, this is far from the case: the whole
story centres around the problem of deciding how to recognise a genuine
``correlated pair.''
As Clauser and Horne clearly state \cite{Clauser/Horne74}, Bell's tests
involving detection of both `$+$' and `$-$' events cannot validly be applied
unless the {\em emission rate of pairs} is known.

The need for the present paper is exemplified by Colin Jack's recent
statement in an article in {\em Physics World} \cite{Jack95}.  He says that
alternatives to the quantum mechanics (QM) explanation are ``even 
more bizarre.''  Have the valid explanations not reached the main 
stream of physics?  If so,
this could be partly because they do not go as far as to say that experiments
prove QM actually {\em wrong}.  Another  reason is likely
to be their specialist, highly mathematical nature.

I have attempted to redress the balance.  It appears to me that the QM
prediction is only approximately correct in certain situations; it
is actually wrong in others.
Data that could have shown the {\em necessity} for
a hidden-variable explanation is likely to have been discarded as being
``not good enough'' for a valid Bell's test.  Also,
the continued use of a biased formula in my view discredits the conclusions
of all the  experiments
covered by the present paper.   This is not intended as a criticism of 
individual experimenters.  Their decisions have been the  results of
their ``acculturation'' (Marshall, Santos, and Selleri's term) in quantum
mechanical ideas.  Neither is it a criticism of the actual experiments.
Alain Aspect's work, as reported in his thesis \cite{Aspect83}, appears to
be excellent.  Valuable opportunities to find out how ``objective reality''
really works, laying the foundations for replacements for QM, are being
wasted, however, if we interpret his results only within the framework of QM
preconceptions.

\begin{figure}
        \centering
        \leavevmode
        \epsfbox{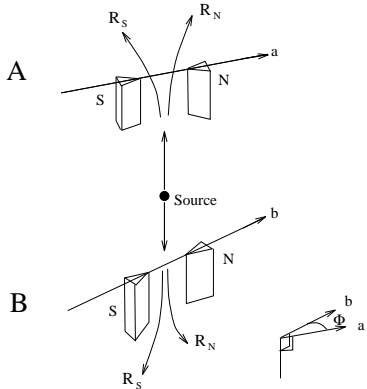}
        \caption{The basic Bohm ``Stern-Gerlach'' experiment.
        $A$ and $B$ label the two halves of the
experiment, S and N are magnets (two of each), and
$R_S$ and $R_N$
detectors.  The settings are defined by the vectors $a$ and $b$, at
relative angle $\phi$.}
        \label{vert_s_g}
\end{figure}

The model that forms the core of the present paper is derived from
consideration of the geometry of David Bohm's version of the EPR experiment
\cite{Bohm51}, with some help from a diagram of Alastair Rae's \cite{Rae81}.   Bohm's idea involved pairs of ``correlated'' particles that move in opposite directions from a common source.  They have opposite ``spins,'' where, under realistic hidden variable theory, spin is the hidden variable, a real-valued vector.  They are deflected towards one or other of the appropriate pair of
detectors when they pass through a non-uniform magnetic field, as shown
in Fig.~\ref{vert_s_g}, which way they are deflected being determined by
their spin.  The QM description of the setup says that spin is a quantised variable that does not exist until it is measured, when it acquires the value $+1$ or $-1$.  The two particles are described by a single ``wave function,'' representing a superposition of states.  The wave function ``collapses'' when we observe one particle, leaving the second particle in a state that is influenced by the {\em setting of the first detector}.  This constitutes the infamous quantum mechanics ``action-at-a-distance'' effect.

The apparatus is shown with the particles moving vertically, rather than the more usual representation with them horizontal, to facilitate comparison with my model.  


This experiment has never, unfortunately, been done in practice, and, indeed, there are theoretical reasons to doubt if it can be done \cite{Franca92-2}.
I assert that, nevertheless, despite slight differences in geometry,
the model based on it provides realistic explanations for a number of actual EPR experiments.

The arrangement of this paper is as follows.  The  first section introduces
the basic analogy (``The Chaotic Ball'').  Next there is discussion of its
implications in relation to Bell's test (``Missing Values in Bell's 
Inequalities'').  We then pass through two stages of ``Linking with Reality,'' moving from hypothetical Stern-Gerlach experiments to a brief glance at actual cascade and parametric down conversion paired-photon experiments.  The fourth section presents a few of the model's predictions, then some ideas are  given for further work and conclusions are drawn.

Further papers are intended, (a) covering the possibility that time variations
could exist in certain EPR experiments and could be a factor in causing
violations of Bell's inequalities; (b) elaborating on the asymmetry
anomaly (inequality of `$+-$' and `$-+$' coincidence counts) 
mentioned in Aspect's thesis and introduced briefly in the ``Linking with Reality'' section; (c) looking at the effect of amplitude correlations between ``photons'' at the point of emission.

\section{The Chaotic Ball}
\begin{figure}
        \centering
        \leavevmode
        \epsfbox{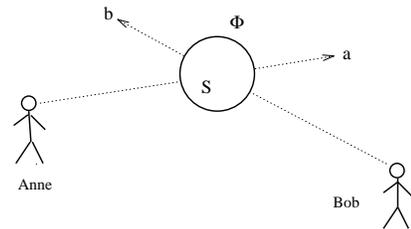}
        \caption{The Chaotic Ball.
        Vectors $a$ and $b$ define directions from assistants to the
        ball.  The angle between them is $\phi$.  The letter $S$ is visible;
        there is an $N$ out of sight opposite to it.}
        \label{anne_bob}
\end{figure}

In this section,
a model is presented that has the same mathematical
properties as an ``EPR'' experiment.
The reader is requested to keep his mind firmly
on the world of every-day experience, and to be patient if the model appears
to be trivial.  Understanding of the ideas covered is an essential
pre-requisite for the remainder of the paper.  No attempt is made to list
all the assumptions behind the model.  There are many.  They are mostly
covered by the statement that everything is as straightforward as it seems.

We have a ball on which just two points are marked, $N$ and $S$, at opposite
poles.  It is fixed at its centre but moving in some chaotic
 fashion so that its orientation at
any given instant is effectively random.  Assistants Anne ($A$) and Bob ($B$)
stand and look at it, Anne in direction~$a$ and Bob in 
direction~$b$, as shown in  Fig.~\ref{anne_bob}.
 In other words,
Anne and Bob are our ``detectors" and their relative angle is $\phi$, the
angle between the vectors $a$ and $b$.  

The assistants each have record sheets with two columns,
 head\-ed $N$ and $S$, and $T$ rows.
At agreed times they put a tick in the appropriate column according as to
whether they see an $N$ or an $S$ facing them.  When the sheet is filled
they get together and fill in another sheet that summarises the ``coincidences".
They put ticks in columns headed $NN$, $NS$, $SN$ and $SS$ as appropriate.
Now they have all they need to estimate
the ``correlation" between their two sets of results.  They hope this will
give them information as to whether they really were looking at the same
ball or whether it was an optical illusion and there were actually two
independent ones or ones whose motion was only slightly correlated.

\begin{figure}
        \centering
        \leavevmode
        \epsfbox{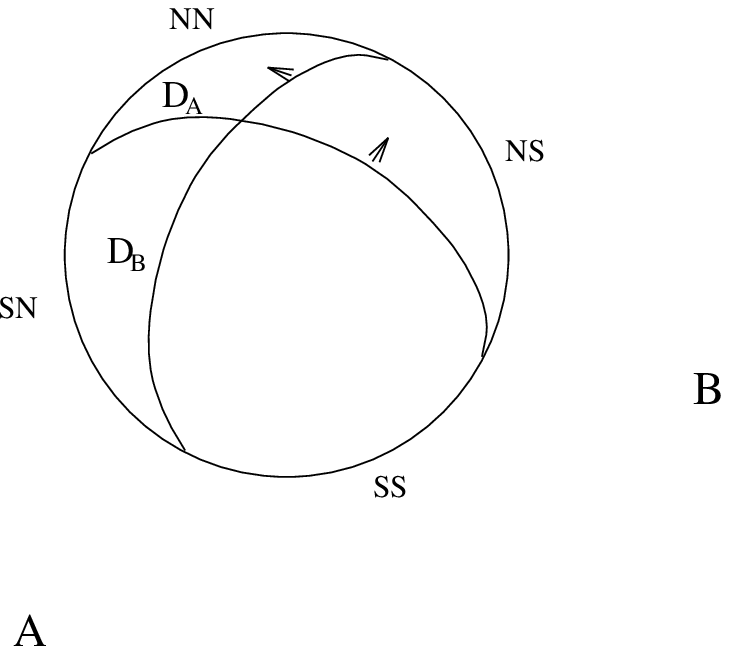}
        \caption{Fixed sphere with ``wire'' discriminators.
                Assistants are abbreviated to $A$ and $B$.
                $SN$, $NN$ etc.\ indicate what the assistants record when the
                $S$ is in that region, the first letter being associated with
                $A$ and the second with $B$.  $D_A$ and $D_B$ are
		discriminators between $N$ and $S$ regions.  Arrows indicate
		direction from $S$ to $N$.}
        \label{spin_sphere}
\end{figure}
\begin{figure}
        \centering
        \leavevmode
        \epsfbox{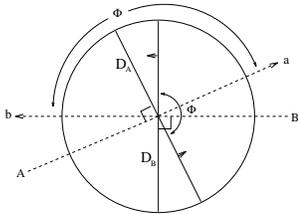}
        \caption{Fixed sphere viewed from above.
        Labelling is as in previous diagram.}
        \label{sphere_top}
\end{figure}

Suppose for the present that there is just the one ball.  The numbers of
ticks in any category should be proportional to the areas between
the circles shown in Figs.~\ref{spin_sphere} and \ref{sphere_top}.

These diagrams can be thought of as representing a fixed ``abstract"
 sphere surrounding our moving ball.   The circles (discriminators, $D_A$ and
$D_B$) dividing the `$+$' ($=N$) regions from the `$-$' ($=S$) regions can be imagined as being made of thin wires, somehow controlled by the positions of the assistants so that their planes keep at right angles to the directions of the assistants.  If the assistants move, so do the circles.  To construct the diagram, we indicate on the ball the classifications corresponding to the `$S$' being in the given position.

It will be seen that the areas between the circles are proportional
to $\phi$ ($NS$ and $SN$ areas) or to $(\pi - \phi)$ ($NN$ and $SS$ areas),
 making the expected values of $NN$ {\em etc.}\ linear functions of 
$\phi$\footnote{In moving from individual observations to joint ones, our
model has assumed ``factorability.''  If we define $p(\lambda,a)$, for
example, to be the probability that Anne sees an $N$ when the ball has the
vector from $S$ to $N$ in the direction $\lambda$, and $p_{NN}(\lambda,a,b)$
to be the joint probability that both Anne and Bob see an $N$, then
$p_{NN}(\lambda,a,b) = p(\lambda,a) p(\lambda,b)$.  John Bell took this
assumption to be effectively a definition of ``locality.''  It is what one
would expect if the two assistants act independently.  In the present instance,
all probabilities are $0$ or $1$ and the relationship is trivial.}.  This is
perhaps easier to see if you visualise dividing up an apple into four pieces using two cuts of a large knife.  The four pieces will be of two different sizes, the size (and the area of skin, which is what we are currently interested in) being proportional to the angle between the cuts.  Notice, incidentally, that it is only the {\em rotational symmetry} of the apple that is needed to make the linear relationship true.  We restrict attention for the present to values of $\phi$ not greater than $\pi$.

We can now evaluate the ``correlation function'' estimated by:
\begin{equation}
	\label{Correlation Function formula}
        \hat{C} = \frac{NN + SS - NS - SN}{T},
\end{equation}
where here $NN$ {\em etc.}\ stand for numbers of ticks.

What is the result of this fascinating study?

Assuming $T$ is large enough for random errors to be negligible, Anne and Bob should find that
\begin{displaymath}
       \hat{C} = \left\{
        \begin{array} {rl}
	+1, & \phi = 0\\
	0,  & \phi = \frac{\pi}{2}\\
	-1, & \phi = \pi.
        \end{array}
        \right.
\end{displaymath}
	
For intermediate values of $\phi$ we get intermediate estimates of $C$, obtaining a linear relationship with $\phi$ as a mathematical consequence of the constancy of the denominator and the linearity of each term of the numerator.

This is all very obvious (so long as diagrams and intuition are admissible!),
and might seem  scarcely worth writing, and yet it has to be said.  It is when the model is modified to cover an imperfection or two that it starts to make some remarkable predictions, a few of which will now be discussed. 

\subsection{Missing Bands}

\begin{figure}
        \centering
        \leavevmode
        \epsfbox{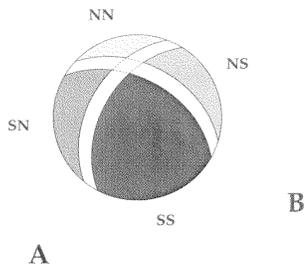}
        \caption{Fixed sphere with missing bands.
                Shaded areas are ``visible'' to both assistants.}
        \label{bands}
\end{figure}

Suppose that Anne and Bob are standing rather close to the ball so that
{\em they can see neither $N$ nor $S$ when the ball is in certain positions}.
These positions will lie in bands determined by the positions of the
assistants.  On our abstract sphere of Fig.~\ref{spin_sphere}, we find that the wire circles that separated our various regions are now bands of finite width.  And look at what happens to our four regions.  They are all reduced, some perhaps vanishing entirely (see Fig.~\ref{bands}).  It is important to realise
that the smaller areas are always affected proportionately more than the larger ones, as we are trimming off strips of fixed width.  Also, the total area ``missing'' varies according to the amount of overlap of the bands.  It is greatest when the bands are at right angles; least when they coincide.

As far as correlation estimates are concerned, {\em when Anne sees nothing Bob's record has to be ignored} and {\em vice versa}.  It is no longer obvious what we should use as our denominator.  Let us do what we would do in practice in an EPR experiment.  Instead of $\hat{C}$, let us use
\begin{equation}
	\label{illegal correlation formula}
	E = \frac{NN + SS - NS - SN}{NN + SS + NS + SN}.
\end{equation}

\begin{figure}
        \centering
        \leavevmode
        \epsfbox{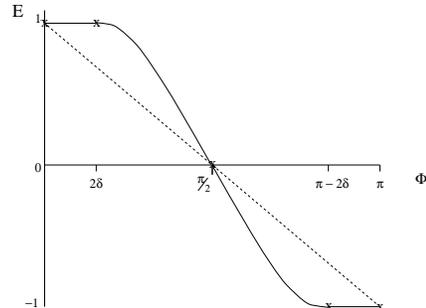}
        \caption{Estimated correlation with missing bands (approximate).
		$E$ is correlation using ``illegal'' formula.
		$\phi$ is angle between assistant directions.
		$\delta$ is half-angle subtended at centre of sphere by
		missing bands.  Points $\times$ are exact.  The curve is drawn
		by eye.}
	\label{est_corr}
\end{figure}

Superficially, this looks a remarkably good substitute.  We get the {\em same} values, on average, as we did before for $\hat{C}$,for $\phi$ equal to 0, $\pi/2$ and $\pi$.  Also, if Anne were to use just the information that figured in the new correlation estimate to estimate her probability of seeing an $N$, she would obtain about $0.5$, {\em as before}.  Of course, if the assistants are standing so close to the ball that each sees only a small disc, there will
be values of $\phi$ such that they {\em never} score a coincidence --- the
discs never overlap.  We {\em cannot} then estimate $C$ using~(\ref{illegal
correlation formula}) as it will be undefined, both numerator and denominator being zero!  Ignoring this possibility for the present, let us look at the complete relationship between correlation and angle.  The straight line of
the basic model (dotted line in Fig.~\ref{est_corr}) is  replaced by a curve, as shown by the solid line.  The points marked with crosses are accurate, corresponding to situations in which missing bands (subtending half-angles $\delta$ at the centre of the sphere), exactly touch, or some other geometrical constraint holds.  The curve is drawn by eye, partly because (as Pearle discovered back in 1970\footnote{The interested reader will find a diagram that helps explain Philip Pearle's paper published by Risco-Delgado \cite{Risco93}.} \cite{Pearle70}) the mathematics required for accurate calculation is not easy and partly because the model in its present form is not intended to give more than qualitative predictions.

Thus, simply by having some missing values, we have produced an estimate
of correlation that is biased for certain values of $\phi$.  The bias is caused by the fact that the missing values are not random.  This is very important, and more detail is given in Appendix~A.  The correlation curve will look approximately sinusoidal, and will be {\em steeper} in places than the original straight line.  Using the correct formula, with $T$ instead of $T_{obs} = NN + SS + NS + SN$, we get a different shape, never steeper than the straight line and never attaining the $+1$ or $-1$ limits.  Changing our assumptions to make the probability of a null result vary gradually rather than switch straight from 0 to 1 ({\em i.e.}\ giving our invisible bands fuzzy edges) will spread the curve out slightly; reducing the degree of correlation between $N$ and $S$ may have a similar effect (a computer simulation would help clarify this).

If we know how many observations are missing, however, we should  be able to correct our estimate.  But what happens in real experiments?  We generally do not {\em know} directly that we have {\em any} of these ``value-dependent'' missing records.  They are confounded with other ``random'' ones.  Also, their possible existence is obscured by the fact that they have {\em no effect} on some of the most interesting estimates, those for angles $0$, $\pi/2$ and $\pi$.  The question is, ``Can we {\em deduce} the existence of value-dependent missing records?'' and the answer is ``Yes, though it may not be easy!'' 
 One would think that it should sometimes be glaringly obvious, when there are large numbers of missing values and hence values of $\phi$ that give hardly any ``genuine coincidences'' (see diagrams in Section~\ref{predictions}),
but a  real experiment with this feature might well be dismissed as having
too poor a ``ratio of genuine to accidental coincidences'' to be useful.   The existence of missing bands is also in fact obvious in less extreme cases from the general shape of the curve, but let us assume that a rival theory attempts to approximate the correlation by a sine curve.

What symptoms can we look for when there are only moderate missing bands,
symptoms that are {\em not} shared by a sine curve?  Now, for sine
curves, the total number of coincidences that we observe, $T_{obs}$,
is constant. But so long as both Anne and Bob have blind areas, $T_{obs}$, can be seen to vary with $\phi$, the greatest contrast being between the values when $\phi$ is $0$ or $\pi$ and that when $\phi$ is $\pi/2$.  Without ``hidden
values,'' {\em i.e.}\ actual shared information, how can this be accounted for?

\subsection{The Effect of Asymmetry}
\begin{figure}
        \centering
        \leavevmode
	\epsfbox{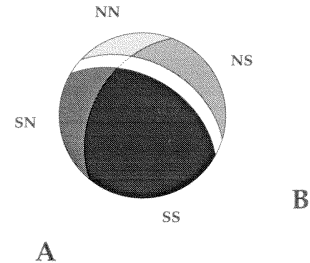}
        \caption{Fixed sphere with one missing band.  
	$A$ represents imperfect
	and $B$ perfect detection. }
        \label{one_band}
\end{figure}

 If one assistant can see the whole ball whilst the other can see only part,
we have an intriguing situation.  $T_{obs}$ is now {\em constant} (see 
Fig.~\ref{one_band}), despite the missing values.  If we cannot look at 
the raw data, the exact shape of the relationship between coincidence numbers and relative angle will in fact contain clues that there are missing values, 
 but they may be difficult to recognise.

Another possible form of asymmetry would occur if, say, the letter `$S$'
were bigger than the `$N$', so that the missing band was smaller on the `$S$'
side.  This  would have the effect of making $T_{obs}$ smaller when the assistants stand opposite each other than when they stand together.  There should, however, be even smaller values when they stand at right angles.

When we have simultaneously both differences between the assistants and differences between `$N$' and `$S$', we get some new possibilities (see 
later discussion of Alain Aspect's experiments, and separate paper).

Serge Caser has published some interesting papers that prove that asymmetry is {\em necessary} to reproduce {\em exact quantum mechanical predictions} \cite{Caser84,Caser87}.  Whilst this may be true, and there are probably asymmetries in all real experiments, the model suggests that, if we confine our attention to looking at just the correlation curve, we might often get sufficiently close to actual experimental results with a symmetrical model for it to be within the margin of experimental error.  This is in agreement with Caser's 1987 paper.

\subsection{Non-Locality?}
We have {\em not}, of course, had any thoughts of ``action at a distance.''
Anne and Bob have not cheated by shouting across the room.  (What we {\em have} done, though, is to use a completely unjustifiable estimate for our ``correlation,'' of which more in later sections).  Yet a quantum theorist, obtaining a curve such as that of Fig.~\ref{est_corr}, might be tempted to say that it was a sine curve and evidence that somehow the system we were observing was managing to obey the mysterious quantum ``collapsing wave function'' rules.

\section{Missing Values in Bell's Inequality}
The chaotic ball model with missing bands illustrates not only that
it is easy to devise real situations that reproduce ``correlation''
predictions that would be difficult to distinguish experimentally
from quantum theory's sine curves, but also that it is easy to violate
several of the ``Bell's Inequalities.''  These are tests that are violated
by quantum mechanics predictions but are supposed not to be by realistic 
``local'' models.  To see how violations arise, all we need do
is to draw the appropriate diagrams and perhaps do a little algebra: the 
inequalities can then be identified with natural geometrical or algebraic
constraints.

\subsection{Simplest Version of Bell's Test}
For the very simplest version, we have to conduct three sub-experi\-ments of the chaotic ball type, with specified relative orientations.  The inequality is
\begin{equation}
\label{first bell's inequality}
        p_{NS}(a, b) + p_{NS}(b, c)
	 \geq p_{NS}(a,c),
\end{equation}
where, for example, $p_{NS}(a, b)$ means the probability of Anne observing an `$N$' and Bob an `$S$' when Anne's direction of observation is given by the angle $a$ and Bob's is  given by $b$, with respect to a fixed laboratory frame.  Each term of inequality~(\ref{first bell's inequality}) corresponds to one sub-experiment.

If there are no missing values (as is {\em assumed} in the derivation of the
inequality --- see, for example, Jim Baggott's book \cite{Baggott92}) then a quick sketch of the appropriate diagrams on the lines of Fig.~\ref{sphere_top} will show that the inequality should be an {\em equality} for some selections of angles.

\begin{figure}
        \centering
        \leavevmode
        \epsfbox{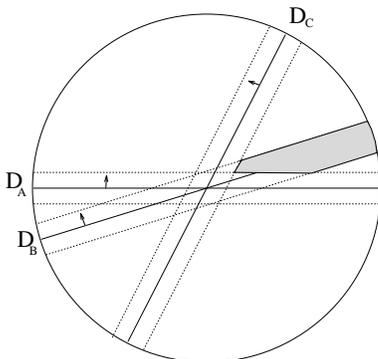}
        \caption{Violation of simple version of Bell's inequality.
                The shaded area is approximately proportional to the Bell's
                discrepancy.
                $D_{A}$, $D_{B}$ and $D_{C}$ are the ``discriminators'' between
                essentially $S$ and $N$ regions.  Arrows indicate the direction
                from $S$ to $N$. }
        \label{bells}
\end{figure}

If there are bands of value-dependent missing records (a special case of ``Variable Detection Probabilities'' (VDPs)), our first question is ``What do we mean by `probability' in this context?''   Do we mean expected proportions out of the number of pairs we started with ($T$), out of the number we observed ($T_{obs}$) or, perhaps, out of the number of ``singles'' that either one of the assistants observed?  Let us assume for the present that we are dealing with the actual Anne-and-Bob experiment, so we know the value of $T$.  We can then use the appropriate ratio ($NS/T$, for example) as our estimate.  Because the denominators in all terms will then be the same, our inequality will depend only on the numerators.  It will be equivalent to 
\begin{equation}
	NS(a,b) + NS(b,c) \geq NS(a,c).
\end{equation}
The difference between the two sides of this expression corresponds to the
shaded area in Fig.~\ref{bells}. {\em The inequality will be violated}.

In the equivalent EPR experiment, of course, it would not be immediately
obvious that there were any (value-dependent) missing records and the standard procedure (not, to be fair, followed by some workers without a certain degree of reluctance) would be to {\em assume there are none}, in line with quantum theory custom.  The probabilities would therefore be estimated by ratios such as $NS/T_{obs}$, where $T_{obs}$ is $NN + SS + NS + SN$ for the current
sub-experiment.  The use of this biased estimate would not, in practice,
affect the behaviour of this version of Bell's test, as the angles used
in actual comparisons would be such that $T_{obs}$ would be almost constant.  The inequality would again be violated.

\subsection{``Standard'' Bell's Inequality}
\label{Standard Bell's Inequality}
The inequality discussed above is not the favoured one.  Its restrictions are too limiting, with no allowance for any stochastic (``fuzzy'') element at all.
The ``standard'' one involves conducting four sub-experiments and is
\begin{equation}
\label{Standard Bell's IQ}
-2 \leq C(a,b) - C(a,b') + C(a',b) + C(a',b') \leq 2,
\end{equation}
where $a$, $a'$, $b$, $b'$ are detector settings and $C$ is officially
\cite{Selleri88}
 estimated by $(NN + SS - NS -SN)/T$, though
{\em in practice $T_{obs}$ is used instead of $T$ } for the denominator.

The erroneous use of $T_{obs}$ can cause violation of~(\ref{Standard Bell's
IQ}) unless the underlying correlation is weak.   Clauser and Horne \cite{Clauser/Horne74} present a proof that demonstrates that the inequality, {\em if correctly applied}, would be true even in the presence of missing bands and/or stochastic elements.  They stress that correct application involves knowing the ``emission rate'' and using this as denominator in the correlation
estimates.

\begin{figure}
        \centering
        \leavevmode
        \epsfbox{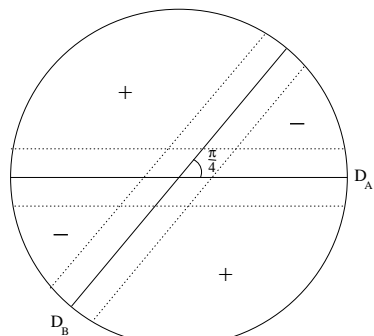}
        \caption{Violation of Bell's inequality due to missing bands.
        $+$ and $-$ symbols indicate the signs with which the regions enter
        into the Bell's test.  See text for further explanation.}
        \label{bells_3}
\end{figure}

It is not quite so easy to illustrate violation diagrammatically as it was
for the simple version.  Let us fix our attention on a particular set of angles that is commonly used in actual Bell's tests: $a = 0$, $a' = \pi/2$,
$b = \pi/4$ and $b' = 3\pi/4$.  If we draw the four diagrams (one is shown in
Fig.~\ref{bells_3}, in which `$+$' and `$-$' signs indicate the sign of the
contribution of the region to the term in question), we find that the value
we require is the sum of four equal contributions, one per diagram.  If we can show that the contribution of any one of them can be more than $1/2$ when we have missing bands, then we shall have demonstrated violation.

The algebra of Appendix~A shows immediately that the contribution {\em will}  exceed $1/2$, for we have $x \propto 3\pi/4$ and $y \propto \pi/4$, so that the contribution if there were no missing bands would be 
\begin{displaymath}
	\frac{x-y}{x+y} = \frac{1}{2}, 
\end{displaymath}
and that with missing bands would be greater than this.

\subsection{``CHSH'' Inequality}
Clauser and Horne state on more general theoretical grounds that inequality~(\ref{Standard Bell's IQ}) {should not be used unless the number of pairs emitted is known}.  They and others devised the third main group of ``Bell's'' inequalities (Clauser, Horne, Shimony, Holt (CHSH) tests), members of which are not so easy to violate as they are unaffected by the type of missing value we have been considering so far.  They are used in some of the most important EPR experiments, notably the Aspect, Dalibard and Roger paired-photon experiment of 1982 \cite{Aspect82}, and  will be covered in a separate paper.

\subsection{Does Violation have any Significance?}
John Bell ``proved" that no hidden-value theory could predict values that violated his inequality.  Whilst this may be almost true of the standard version {\em if applied correctly, using $T$}, the forgoing has demonstrated that it is far from true as used in practice.  Violation is a natural consequence of the existence of missing bands or, more generally, VDPs.  It is associated with the roughly sinusoidal relationship between estimated ``correlation'' (using the illegal equation~(\ref{illegal correlation formula})) and angle that will be produced whenever detection probabilities are  low, unless every signal has an {\em equally low} probability of detection.
Contrary to widely-held beliefs \cite{Aspect83,Clauser/Shimony78}, low detection efficiencies can produce {\em high} ``correlations'' when
estimated in this way\footnote{A referee has kindly drawn my attention to a
paper by Greenberger \cite{Greenberger95} reporting large correlations for
a three-particle model that has logical similarities with the chaotic ball.}.
Violation of the ``standard'' inequality is thus of no significance
whatsoever.

\subsection{How can we distinguish between Quantum Theory and Hidden
Value Theories?}
There are some respects in which we can hope to see that the coincidence
curves produced in EPR experiments are not truly sinusoidal (see examples in Section~\ref{predictions}), and these are what we need to concentrate on once we start applying our method, as we want to bring out the contrast with QT's predictions.

An important difference between our predicted results and those based on
sine curves is that the total number of coincidences, $T_{obs}$, may vary with angle to  give a minimum at $\phi = \pi/2$ ($\pi/4$ for paired-photon polarisation experiments).  We do not expect to see large variations, as those
with gradually varying detection probabilities would not be expected to
be as dramatic as those with hard-edged missing bands.  QM's sine curves,
depending on $\phi$ through a factor of $cos(\phi)$, can produce minima at $\phi = 0$ or $\phi = \pi$, in asymmetric cases \cite{Lepore/Selleri90}, but never at $\phi = \pi/2$.  {\em A minimum near $\pi/2$, therefore, constitutes evidence of the existence of hidden variables}.  It shows that there are missing records whose number varies systematically with angle.  In the light
of the chaotic ball results, we can say that the {\em obvious} cause of
the variation is that the probability of detection varies according to
the value of a hidden variable.  So the latter must exist.

There are numerous other differences.  Some that occur in asymmetric
situations hold promise of finally proving hidden variable theories superior.

Studies of coincidence curves should ideally be accompanied by direct
observation of how the ``singles rate'' varies with the hidden variable.
Assuming this hidden variable to be polarisation (see below), I do not agree with the statements in Clauser et al. 
\cite{Clauser/Horne/Shimony/Holt69} that attempts at measuring
these relationships are ``irrelevant,'' or that ``highly pathological
detectors'' are needed to convert hidden-variables predictions to anything like the QM ones\footnote{Since writing the main body of this paper, the work of Barry Gilbert and Sue Sulcs (Telstra Research Laboratories, 770 Blackburn Road, Clayton Vic 3168, Australia) has come to my attention.  They have produced
a simple computer simulation \cite{Gilbert/Sulcs96} of a radio analogue 
of an EPR experiment.  This includes what in my opinion is an excellent modelling of the detection process, with {\em noise} playing a critical role.  It can reproduce QM predictions to reasonable accuracy and is not in any sense ``pathological.''  Note, incidentally, that the Bell's test they use is the one I refer to as the CHSH one, to be covered in my next paper, not the ``standard'' one discussed here.}.  It is true, as is well known, that we cannot produce an ensemble of signals all with exactly the same polarisation, but if we were to produce one that was approximately of this nature, and if this showed variation of detection probability that was not {\em exactly} sinusoidal, then would this not in itself constitute evidence in favour of hidden variables?

\section{Linking with Reality}
The chaotic ball analogy relates most directly to the original {\em gedanken} experiment, which Bohm translated into a Stern-Gerlach ``spin'' experiment (see introduction).  The principles --- the manner in which variable detection probabilities influence the results --- apply equally to many different EPR variations.  The whole class of models, embodying the kind of realism exemplified by the analogy, will be termed in this paper ``LRMs'': local realistic models.

For ``spin'' experiments, the moving ball of our analogy represents, of course, our source of particles.  The fixed sphere surrounding it is an abstraction representing not the physical geometry of the experiment but that of the set of possible hidden values as they exist at the source.   The model itself comprises this fixed sphere with the effects of the measuring apparatus ``projected back'' onto it.  

The net result is as if the particles undergo no vertical motion and the assistants (one standing behind the `$S$' magnet of each pair) are able to look at a single particle situated at the source.  I have simplified things by modelling identical, not opposite, spins.  Mathematically, the difference is trivial.  If it were ever necessary to convert to the ``opposite spin'' situation, one could just interchange the `$S$' and `$N$' labels on one pair of magnets.

\begin{figure}
        \centering
        \leavevmode
        \epsfbox{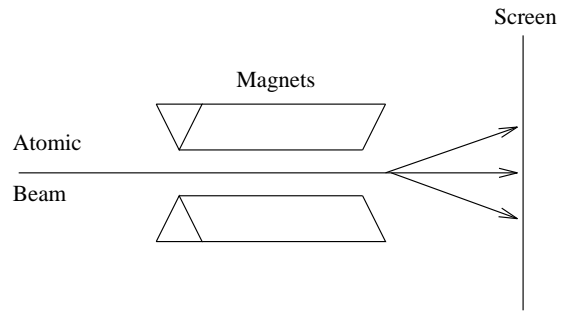}
        \caption{A general Stern-Gerlach apparatus}
        \label{stern_gerlach}
\end{figure}

Suppose that we conduct our experiment, and the observations produce a correlation curve that might be the sine curve predicted by QM, but might equally well be the curve shown in Fig.~\ref{est_corr}, arising from missing bands.  Unless our correlation, by the time it is measured, is weak, we find that Bell's inequality is violated.  Knowing how LRMs work, and given the nature of Stern-Gerlach apparatus, the quantum theory ``explanation'' seems ridiculous.  If, say, particles with spin almost exactly parallel to the apparatus setting are deflected very strongly, so that they perhaps hit one of the magnets, this will be represented on our abstract model as missing ``caps.''  If those with weak components in the required direction pass between the detectors, then these will correspond to one of our Anne-and-Bob-type missing bands.  And if we look at an alternative diagram of a Stern-Gerlach apparatus we see that, even within QM, it might have been wrong to assume
that our source was producing the particular quantum state that gives just
 `$+$' or `$-$' spins.  Fig.~\ref{stern_gerlach} is based on one in Bohm's 1951 text book, page 326.

But this is purely hypothetical.  There have been  no (credible) experiments involving particle spin.  Perhaps this is because real sources cannot be made to produce uniform distributions of spins.  This would present no special problem for our LRM, but probably insuperable difficulties for QM.

The fact of the matter is that the chaotic ball as presented may have no direct applications.  The principles involved, however, will clearly carry over to a large set of related situations.  

\subsection{Bell's ``Standard'' Inequality, Actual Experiments and some
Anomalies}
When talking of real experiments, it is inevitable that interpretation in the light of new theory will involve criticism of existing interpretations.  I should like to remind the reader, therefore, of my earlier remarks: I do not
intend to suggest either that the experiments discussed were conducted in any
but the most exemplary fashion, or that they were in any way unique.  They were conducted and interpreted following accepted customs.

A good illustration of the (mis-)application of Bell's ``standard'' inequality is to be found in Aspect, Grangier and Roger's paper \cite{Aspect82-0}.  The authors carried out several tests that they thought would ensure the validity of Bell's test, but our model shows that they were not sufficient.  The experiment was not symmetrical with respect either to the two photons or to the detection of `$+$' and `$-$' events.  The imbalances are fully discussed in Aspect's thesis \cite{Aspect83}.

The hidden variable  is ``polarisation'' \cite{Shurcliff/Ballard64}, and the
geometry can probably be represented satisfactorily by the perimeter of a circle, divided into four sectors.  Values falling in two opposite sectors correspond to one ``polarisation direction'' and those in the other two to the perpendicular direction.  The missing bands of our analogy would be replaced by missing arcs between sectors\footnote{If light is considered to be composed of
``particles,'' then it is difficult if not impossible to envisage missing (or, double) arcs.  If, on the other hand, it is recognised that experiments such as Grangier, Roger and Aspect's of 1986 \cite{Grangier/Roger/Aspect86} can be explained more satisfactorily by purely wave theories such as those of
Marshall and Santos or Gilbert and Sulcs \cite{Gilbert/Sulcs96,Marshall/Santos87}, then there is no problem.  In
such a theory, individual ``photons'' possess amplitudes.  Note also that there is no doubt in all real experiments that there are {\em some} missing records: efficiencies of detection in the experiment in question are only around 3\% for one ``photon,'' 10\% for the other.}.   (There is no {\em a priori} reason why there should not for certain parameter settings sometimes be {\em double} bands/arcs rather than missing ones.  Under a plausible classical theory, this might be expected when  signal amplitudes are high or detector thresholds are set low.)

Preliminary re-analysis of the published results, which violate Bell's inequality by several standard deviations, suggests that we can explain the violation by assuming a very simple model with missing arcs of half-angle about $\pi/15$ associated with one of the two-channel polarisers, the other being assumed to show no dependency of total output on exact polarisation angle.  This explanation would not conflict with Aspect's statement that he found $T_{obs}$ to be constant.  

Alternatively, one might consider challenging the constancy of $T_{obs}$.  In his thesis, Aspect mentions that it {\em does} vary slightly, though not significantly.  The fact that he feels the variation deserves mention reflects that fact that he knows that, if significant, it would invalidate his Bell test.  Could it be also that the same pattern of variation with $\phi$ shows up on several occasions?  Aspect does not tell us, and, rather than increase replication so as to investigate the effect with greater precision, he adjusts
parameters (primarily the detector thresholds) so as to make the variations
as small as possible.  I shall be explaining in a later paper how negative amplitude correlations may be present and may be unwittingly exploited to serve this end, whilst still producing violation of the ``standard'' Bell test.

Two further ``anomalies'' reported in Aspect's thesis perhaps deserve mention.
They are both perfectly natural under the theory I am presenting here.

Firstly, there is slight evidence that the sum of the `$+$' and `$-$' probabilities varies with polarisation angle.  As is well known (see earlier), this sum cannot be measured directly as one cannot produce an ensemble of signals all with exactly the same polarisation.  Nevertheless, Aspect attempts to deduce from a supplementary experiment an upper limit on its variation.  He
derives a modified Bell's test to correct for it.  It is difficult to follow his reasoning here: would it not make more sense to accept that he is seeing evidence of VDPs at work?  Quite apart from this objection, I would question the assumption he makes, that the value measured in his supplementary experiment is not exceeded by some subpopulations of signals.

Secondly, despite his best efforts in adjusting the apparatus settings to achieve effective equality between the two channels of each polariser, he sometimes finds certain quantities unequal that QM would predict to be equal.  For example the number of `$+-$' coincidences differs slightly from the number of `$-+$' ones for certain values of $\phi$.  Using a realistic model that allowed for different patterns of detection probabilities, as functions of polarisation angle, for each of the four routes through the apparatus, this would be completely natural.  This subject will be elaborated in a later paper.

Rarity and Tapster's 1990 experiment \cite{Rarity/Tapster90} is another example of the application of the standard Bell's test, with the hidden variable this time being phase.  The authors do not report in the paper whether or not $T_{obs}$ is constant, though I gather (personal communication) that it is so, within the accuracy of the experiment.  They are prepared to accept two assumptions (amounting to ``factorability'' and an assumption about instrument settings ``redistributing'' the `$+$' and `$-$' results, for fixed hidden variable, but not altering the total), quoting Grangier, Potasek and Yurke \cite{Grangier88} as their authority.  As I shall show in my next paper, the first assumption is not so obviously correct as is generally thought; the second, in the light of my chaotic ball model, cannot be justified.  To test it, even approximately, would require considerable ingenuity.

\section{Predictions of the LRM}
\label{predictions}

The models can produce predictions of the standard test statistics, but much
information has been lost if we do not also look at the raw data.  It is
recommended that experimenters look at the whole pattern of variation of raw coincidence rates with relative detector angle.  There may sometimes be a case for standardising by dividing by an estimate of the emission rate for pairs, but, unless the aim is {\em  solely} to compare results with published work, we should never be dividing by $T_{obs}$.

\begin{figure}
        \centering
        \leavevmode
        \epsfbox{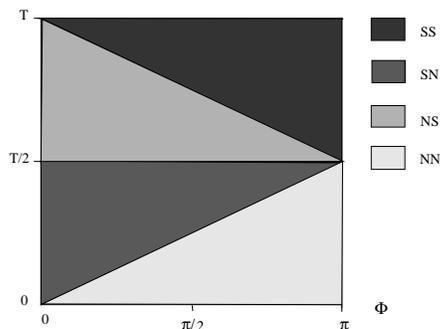}
        \caption{LRM prediction for perfect correlation and detection.
		$T$ is the total number of pairs, $\phi$ the angle between
		the directions of the assistants, and $SS$, $SN$ etc. indicate
		numbers of records expected in each category.}
        \label{lrm_perfect}
\end{figure}

\begin{figure}
        \centering
        \leavevmode
        \epsfbox{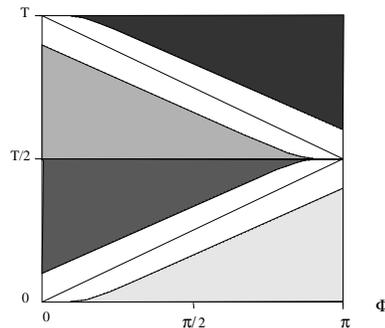}
        \caption{LRM prediction with modest missing bands, perfect correlation}
        \label{lrm_usual}
\end{figure}

\begin{figure}
        \centering
        \leavevmode
        \epsfbox{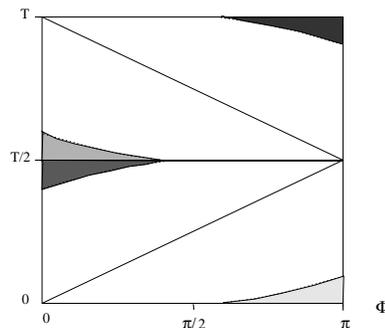}
        \caption{LRM prediction with large missing bands, perfect correlation}
        \label{lrm_extreme}
\end{figure}

\begin{figure}
        \centering
        \leavevmode
        \epsfbox{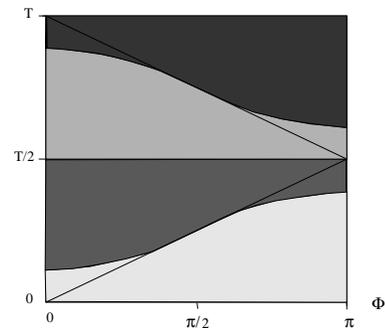}
        \caption{LRM prediction with imperfect correlation, no missing values}
        \label{lrm_lower_corr}
\end{figure}

The graphs (Figs.~\ref{lrm_perfect} to \ref{lrm_lower_corr}) give a general
 idea of the effects to be expected under certain ideal symmetrical conditions.   They can be extended to cover wider ranges of $\phi$ by repeating the patterns,
alternately as shown and their mirror images.  A spherical hidden variable space is assumed: the qualitative predictions for other spaces would be the same.  Lower dimensions officially produce straighter lines, but, in practice,
``fuzziness'' would always be expected to introduce curvature.

The variable graphed is raw coincidence rate (see above).

In practice, of course, $T$ is never known.  The graphs could be produced by simulation models, but the real-life equivalents would have to be drawn with estimates for the white (missing) areas.  

Experimenters will hopefully fully understand the model as it applies to their apparatus, and be able to create their own predictions (with full allowance for that most important property, asymmetry), perhaps with the aid of computer simulation.  They should be able to judge for themselves where the greatest contrast with QM is to be found.  As a general rule,  the {\em shape} of the white area is important: if it is of constant width this means $T_{obs}$ is constant, which means that we need to look at other features to discriminate.  If it is widest for $\phi = \pi/2$ we need look no further for evidence that QM is wrong.

\begin{figure}
        \centering
        \leavevmode
        \epsfbox{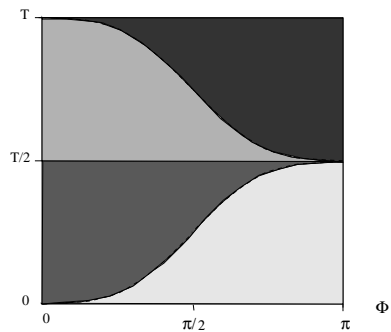}
        \caption{QM Prediction}
        \label{qt_perfect}
\end{figure}

The (symmetric) quantum theory prediction is also graphed for comparison 
(Fig.~\ref{qt_perfect}).  This prediction can vary in vertical scale only, being padded out evenly with white space if detectors are not perfect.

\section{Further Work}
We need to look at the extreme cases.  EPR work so far has concentrated on one end of the spectrum --- ``perfection'' --- aiming at better and better photomultipliers \cite{Kleinknecht86} {\em etc.}.  It is now clear that it should be easier to discriminate between LRMs and QM by looking at the other end --- ``poor'' experiments, using inefficient detectors.  We need to recognise explicitly asymmetries that have, in practice, been ignored.  They have been assumed not to exist if the apparatus has been adjusted so that the
total counts were equal.  We also need to concentrate more on the contrast between the parallel and perpendicular detector settings, rather than the intermediate angles that are ``best'' for Bell's tests.

One interesting way of producing a ``poor'' experiment is to place the detectors at great distances.  It is left as an exercise for the reader to predict what happens to the correlation ``wave function'' as distance is increased.  Another way is to adjust photomultipier voltage and/or
``discriminator'' threshold.

The LRM can make straightforward predictions in areas where QM experiences
difficulty.  For example, it can model non-uniform sources.  Such experiments will not give comparisons with QM, but may be useful in their own right.

We must not forget that there could be other factors at work.  Could
{\em timing variations}, the main topic covered in my second paper, have
 some effect?

\section{Conclusion}
Thus, in my opinion, we can now explain how one area of QM achieves its ``success'' --- the formula for correlated particles gives, in certain cases, approximately the correct prediction for the behaviour of a {\em biased estimate of the correlation}.  This estimate is {\em not} covered by Bell's inequalities, so violation of the inequalities is of no significance.  

QM will be found to give wrong predictions when the ``quantum efficiency'' of the measuring apparatus is either too high {\em or too low}.  The LRM presented here can (aided, perhaps, by computer simulation, such as the one by Gilbert and Sulcs mentioned in a footnote) make valid predictions in areas where
QM is at a loss.  It may, for example, be able to explain a few anomalies
reported by Aspect.

From a theoretical point of view, the important conclusion is that the
``hidden variables'' that cause the observed correlations {\em exist}.  In other words, variables whose ``quantisation'' gave rise to the name of the theory are {\em not quantised at all}.  EPR experiments enable us, for the first time in history, to (as it were) glimpse these variables from two angles at once.

The counterintuitive ideas of quantum theory can be rejected, in this context at least.  We can once again believe in the existence of ``objective reality,''
and this means that we can be certain that those parts of quantum theory that correspond to reality will be found to be justifiable by alternative reasoning.

\appendix

\section{Bias Caused by VDPs}
\label{algebra}

Consider the estimate used in practice for ``correlation'':
\begin{equation}
        E = \frac{NN + SS - NS - SN}{NN + SS + NS + SN},
\end{equation}
or, putting $NN = SS = x$ and $NS = SN = y$, assuming perfect detectors and
rotational symmetry, and taking the expression as representing the limiting case as the numbers of observations tend to infinity:
\begin{equation}
        E = \frac{x-y}{x+y},
\end{equation}
after cancelling 2s.

Now look at a system with fairly narrow missing bands (a special case of
VDPs), as shown in Fig.~\ref{bands}.  Consider values of $\phi = (b-a)$
not too near $0$ or $\pi$.  Then we have, approximately, a {\em constant}
amount, $\Delta$, say, {\em subtracted} from each area.  Our estimate becomes

\begin{equation}
\frac{(x - \Delta) - (y - \Delta)}
{(x - \Delta) + (y - \Delta)} =
\frac{x - y}{x + y - 2\Delta}.
\end{equation}

This is clearly numerically greater than
\begin{equation}
\frac{x - y}{x + y},
\end{equation}
whenever $\Delta > 0$.  The presence of missing bands (or, to a lesser extent,
more gradually-varying VDPs) {\em exaggerates} the correlation.

Under QM assumptions, this does not happen.  QM tacitly implies that if there
are missing detections they are {\em random}.  This means that the effect is
to multiply $x$ and $y$ by a constant, not subtract one.  If we assume
``detector efficiency'' is $\eta$, then we find our estimate for $E$ has expectation
\begin{equation}
\frac{\eta^2 x - \eta^2y}{\eta^2x + \eta^2y}
= \frac{x - y}{x + y}.
\end{equation}


\begin{thebibliography}{00}

\bibitem [*] {Ref}
Published as: Foundations of Physics Letters 9, 357-382 (1996).

\bibitem [**] {CHT/email+}
Email: c.h.thompson@dial.pipex.com\\
Web Site: http://www.aber.ac.uk/$\tilde{~}$cat

\bibitem{Aspect83} Aspect, A.,
                ``Trois tests exp\'{e}rimentaux des in\'{e}galit\'{e}s de
                {B}ell
                par mesure de corr\'{e}lation de polarisation de photons,''
                Ph.D. thesis, Universit\'{e} de Paris-Sud,
                Centre D'Orsay (1983).
\bibitem{Aspect82} Aspect, A., Dalibard, J., and Roger, G.,
                ``Experimental test of {B}ell's inequalities using
                time-varying analyzers,'' {\em Phys. Rev. Lett.} {\bf 49},
                1804-1807 (1982).
\bibitem{Aspect82-0} Aspect, A., Grangier, P., and Roger, G., ``Experimental
	realization of {E}instein-{P}odolsky-{R}osen-{B}ohm
                {\em gedanken} experiment: A new violation of {B}ell's
                inequalities,'' {\em Phys. Rev. Lett.} {\bf 49},
		91-94 (1982).
\bibitem{Baggott92} Baggott, J., {\em The {M}eaning of {Q}uantum {T}heory}
		(Oxford University Press, Oxford, 1992).
\bibitem{Bell64} Bell, J. S., {\em Physics} {\bf 1}, 195 (1964).
\bibitem{Bohm51} Bohm, D., {\em Quantum {M}echanics}
		(Prentice-Hall, Englewood Cliffs, 1951).
\bibitem{Caser84} Caser, S., ``Objective local theories and the symmetry
		between analysers,'' {\em Phys. Lett. A} {\bf 102}, 
		152-8 (1984).
\bibitem{Caser87} Caser, S., ``Quantum mechanics as the limit of a 
		symmetric local theory,'' {\em Phys. Lett. A} {\bf 121},
		331-333 (1987).
\bibitem{Clauser/Horne74} Clauser, J.~F., and Horne, M.~A.,
		{\em Phys. Rev. D} {\bf 10}, 526-35 (1974).
\bibitem{Clauser/Horne/Shimony/Holt69}
                Clauser, J.~F., Horne, M.~A., Shimony, A.,
                and Holt,~R.~A.,  {\em Phys. Rev. Lett.} {\bf 23},
                880-884 (1969), footnote 8.
\bibitem{Clauser/Shimony78} Clauser, J.~F., and Shimony, A., {\em Rep. Prog.
		Phys.} {\bf 41}, 1881-1927 (1978).
\bibitem{Einstein/Podolsky/Rosen35} Einstein,~A., Podolsky,~B., and Rosen,~N.,
		``Can quantum-mechanical description of physical reality
                be considered complete?'' {\em Phys. Rev.} {\bf 47}, 777-780
		(1935).
\bibitem{Franca92-2} Fran\c{c}a,~H.~M., Marshall,~T.~W., Santos,~E., and
		Watson,~E.~J., ``Possible interference effect in 
		{S}tern-{G}erlach phenomenon,'' {\em Phys. Rev. A} {\bf 46},
		2265-2270 (1992).
\bibitem{Gilbert/Sulcs96} Gilbert, B.~C., and Sulcs, S., 
		``An exception to {B}ell's theorem,'' submitted to {\em Eur.
		J. Phys.}, (1996).
\bibitem{Grangier88} Grangier, P., Potasek, M.~J., and Yurke, B.,
                ``Probing the phase coherence of parametrically generated
                photon pairs: A new test of {B}ell's inequalities,''
                {\em Phys. Rev. A} {\bf 38}, 3132-3135 (1988).
\bibitem{Grangier/Roger/Aspect86} Grangier, P., Roger, G., and Aspect, A.,
		`` Experimental evidence for a photon anticorrelation effect
                on a beam splitter: a new light on single-photon 
		interferences,''
		{\em Eur. Lett.} {\bf 1}, 173-179 (1986).
\bibitem{Greenberger95} Greenberger,  D.~M.,
		``Two-particle versus three-particle {EPR} experiments,''
		{\em Ann. N. Y. Acad. Sci.} {\bf 755}, 585-599 (1995).
\bibitem{Jack95} Jack, C., ``{S}herlock {H}olmes investigates the {EPR}
		paradox,'' {\em Physics World}, April (1995).
\bibitem{Kleinknecht86} Kleinknecht, K., {\em Detectors for {P}article 
		{R}adiation} (Cambridge University Press, Cambridge, 1986).
\bibitem{Lepore/Selleri90} Lepore, V.~L., and Selleri, F.,
		``Do performed optical tests disprove local realism?''
		{\em Found. Phys. Lett.} {\bf 3}, 203-220 (1990).
\bibitem{Marshall/Santos87} Marshall, T.~W., and Santos, E.,
		``Comment on `{E}xperimental evidence for a  photon
                anticorrelation effect on a beam splitter: a new light on
                single-photon interferences'\ '': {\em Eur. Lett.} {\bf 3},
		293-6 (1987).
\bibitem{Marshall/Santos/Selleri83} Marshall, T.~W., Santos, E., and 
		Selleri, F.,
		``Local realism has not been refuted by atomic-cascade
                experiments,'' {\em Phys. Lett. A} {\bf 98}, 5-9 (1983).
\bibitem{Pascazio88} Pascazio, S., ``Variable detection probability models for
                {E}instein-{P}odolsky-{R}osen-type experiments,''
		{\em Quantum Mechanics versus Local Realism: The
                Einstein-Podolsky-Rosen Paradox}, F. Selleri, ed.
		(Plenum, New York, 1988), p391.
\bibitem{Pearle70} Pearle, P., ``Hidden-variable example based upon data 
		rejection,'' {\em Phys. Rev. D} {\bf 2}, 1418-25 (1970).
\bibitem{Rae81} Rae, A., {\em Quantum {M}echanics} (McGraw-Hill, New York,
		1981).
\bibitem{Rarity/Tapster90} Rarity, J.~G., and Tapster, P.~R.,
		``Experimental violation of {B}ell's inequality based on
                phase and momentum,'' {\em Phys. Rev. Lett.} {\bf 64},
		2495-2498 (1990).
\bibitem{Risco93} Risco-Delgado, R.,
		``The variable detection approach: a wave particle model,''
		{\em Found. Phys. Lett.} {\bf 6}, 399-428 (1993).
\bibitem{Selleri88} Selleri, F.,
		{\em Quantum Mechanics Versus Local Realism:  {T}he
                {E}instein-{P}odolsky-{R}osen {P}aradox}
		(Plenum, New York, 1988), p19.

\bibitem{Shurcliff/Ballard64} Shurcliff, W.~A., and Ballard, S.~S.,
		{\em Polarized {L}ight} (Van Nostrand, New York, 1964).
\end{thebibliography}
\end{document}